# Broadband Single-Mode Hybrid Photonic Crystal Waveguides for Terahertz Integration on a Chip

*Haisu Li\*, Mei Xian Low, Rajour Tanyi Ako, Madhu Bhaskaran, Sharath Sriram, Withawat Withayachumnankul, Boris T. Kuhlmey, and Shaghik Atakaramians\**

Dr. H. Li
Key Laboratory of All Optical Network and Advanced Telecommunication Network of EMC, Institute of Lightwave Technology, Beijing Jiaotong University, Beijing 100044, China
E-mail: lihaisu@bjtu.edu.cn

Dr. H. Li, Dr. S. Atakaramians
School of Electrical Engineering and Telecommunications, UNSW Sydney, New South Wales 2052, Australia
E-mail: s.atakaramians@unsw.edu.au

M. X. Low, R. T. Ako, Prof. M. Bhaskaran, Prof. S. Sriram
Functional Materials and Microsystems Research Group and the Micro Nano Research Facility, RMIT University, Melbourne, Victoria 3000, Australia

Dr. W. Withayachumnankul
School of Electrical and Electronic Engineering, the University of Adelaide, South Australia 5005, Australia

A/Prof. B. T. Kuhlmey
The University of Sydney, School of Physics and Institute of Photonics and Optical Science (IPOS), Camperdown, New South Wales 2006, Australia

Broadband, low-loss, low-dispersion propagation of terahertz pulses in compact waveguide chips is indispensable for terahertz integration. Conventional two-dimensional photonic crystals (PCs) based terahertz waveguides are either all-metallic or all-dielectric, having either high propagation losses due to the Ohmic loss of metal, or a narrow transmission bandwidth restricted by the range of single-mode operation in a frequency range defined by the PC bandgap, respectively. To address this problem, a hybrid (metal/dielectric) terahertz waveguide chip is developed, where the guided mode is completely confined by parallel gold plates and silicon PCs in vertical and lateral directions, respectively. A unique multi-wafer silicon-based fabrication process, including gold–silicon eutectic bonding, micro-patterning, and Bosch silicon etching, is employed to achieve the substrate-free hybrid structure. Theoretical and experimental investigations demonstrate that the hybrid waveguide supports a single-mode transmission covering 0.367-0.411 THz (bandwidth of 44 GHz, over twice wider than that of all-silicon PC waveguides) with low loss (below 0.05 dB mm$^{-1}$) and low group velocity dispersion (from -8.4 ps THz$^{-1}$ mm$^{-1}$ to -0.8 ps THz$^{-1}$ mm$^{-1}$). This work enables more compact, wideband terahertz waveguides and auxiliary functional components that are integratable in chips towards ultra-high-density integrated terahertz devices in particular in the field of wireless communications.

**1. Introduction**

The recent decade has witnessed a burgeoning interest in using terahertz radiation (loosely defined from 0.1 THz to 10 THz) as a promising candidate for numerous scientific and technical applications including communications, sensing, and imaging.[1, 2] Nonetheless, optics-based terahertz systems, which usually incorporate many optical devices and complicated optical paths, are bulky and require precise manual alignment. Compact and robust terahertz waveguides with low loss and low dispersion provide a superior alternative to free-space propagation, which are in vital demand for stable, portable, integrated terahertz systems.[3, 4] To date, various terahertz waveguide solutions based on technologies from both electronics and photonics have been proposed, generally including fiber-like waveguides[5-12] and planar chip-like waveguides.[13-23] In contrast to terahertz fibers, terahertz waveguide chips are flat enough to be integrated into larger devices and able to append functionalities inside the waveguide chips.

One of the simplest planar terahertz waveguides is the metallic parallel-plate waveguide (PPWG), which can support a strict single-mode operation [the fundamental mode is the transverse electromagnetic (TEM) mode].[13] Even though the metallic surface causes Ohmic losses, the TEM mode in PPWG has merits of no cut-off frequency, negligible group velocity dispersion (GVD), as well as relatively low propagation loss (most terahertz power is propagating in the air gap region between parallel plates – dry air has very low absorption in the terahertz range).[13, 24] In addition to Ohmic losses, the TEM mode of PPWG is only confined in one direction, and the lack of confinement in the lateral direction causes additional divergence loss. To prevent the lateral leakage, Bingham and Grischkowsky proposed an all-metallic photonic crystal (PC) waveguide comprising metallic cylinders in a square lattice in 2008.[16] This metallic PC waveguide exhibits a passband covering 0.6-1.1 THz, but higher losses (0.1-0.19 dB mm$^{-1}$) compared to that of PPWG. PCs are periodically modulated structures, being able to prohibit or allow light propagation of specific polarization over certain frequency ranges thanks to the existence of the photonic bandgap (PBG).[25] The same group had also demonstrated functionalities such as high-Q cavities by removing one or more designated cylinders in the all-metallic PC waveguide to achieve linewidths down to 13 GHz.[26] In addition, metal-based transmission lines (for example, coplanar waveguide and microstrip line) have been proposed to propagate terahertz radiation covering over 100 GHz bandwidth.[20, 21] Nonetheless, the propagation losses are of the order of several dB cm$^{-1}$ because of strong mode confinement and involvement of a lossy dielectric substrate, and in particular, the losses increase significantly at high frequencies.

In parallel to the hollow-core all-metallic planar terahertz waveguides, efforts have been made to develop solid-core all-dielectric planar waveguides, mostly using high-resistivity silicon since it is a low-loss material with very low dispersion at terahertz frequencies. In 2006, the first micromachined silicon rectangular terahertz waveguide was fabricated.[15] To reduce the fabrication complexity and improve the dimensional accuracy, several low-loss terahertz waveguides fabricated using silicon-on-glass technology were reported, such as dielectric microstrip lines.[17, 18, 22] However, the relatively high refractive index of silicon narrows the single-mode operation bandwidth,[18, 27] or the waveguide operates in a multimode frequency range.[17] Benefiting from the low-loss properties of silicon in the terahertz range, silicon-based PC slabs (circular air holes in a triangular lattice) have also been proposed for guiding terahertz radiation in 2015 and 2017.[14, 19] Such PC waveguides can offer low propagation losses (below 0.01 dB mm$^{-1}$) in a very narrow frequency range (0.326-0.331 THz),[19] i.e., their bandwidth is two orders of magnitude narrower than that of their metallic counterparts.[16] As proposed recently, the bandwidth of silicon PC waveguides can be enhanced by modification of the rows adjacent to the silicon channel.[28] One major problem of silicon-based PC waveguides is the existence of high order modes even in the first PC bandgap, which would yield unexpected modal coupling and modal dispersion when waveguide-based functionalities such as bends, splitters, and resonators are introduced.[29-32] In other words, the high refractive index of the silicon channel further limits the useful operating (single-mode) transmission bandwidth of the waveguide.

For ultra-wideband short-range terahertz communications systems, low-loss and broadband transmission of terahertz wave is desirable. However, the aforementioned all-metallic PC waveguides are very lossy and the silicon PC waveguides have very limited single-mode bandwidth. In order to balance the narrow bandwidth and propagation loss of terahertz waveguides, we introduce a unique approach inspired by the well-known field of PCs. It is common knowledge that two-dimensional (2D) PC rods in a square lattice have a wider maximum PBG than that of the 2D PC holes in a triangular lattice.[25] However, when the 2D periodic structure becomes three-dimensional (3D; i.e., limited height) the majority of guided modes in the air defect are above the light line and radiated into free space. To overcome this concern and exploit the full single-mode operating bandwidth, we propose a hybrid structure combining metallic PPWG and dielectric rod-type PCs, resulting in a distinctive set of terahertz waveguides and devices with the potential for integration inside a single chip-like structure. The idea of using the hybrid structure relies on the fact that the polarization of the bandgap of the square-latticed rod-type PCs is the same as the polarization of the TEM mode in PPWG, leading to a complete confinement in both directions. Our proposed approach utilizes the PC to provide the lateral confinement in PPWG that is otherwise only achievable using bulk optics in between the plates.[5] Here, we develop a unique silicon microfabrication process to realize the hybrid structure, resulting in compact, substrate-free terahertz chips. Our results demonstrate that the hybrid waveguides can indeed offer a wide single-mode, low-dispersion operating regime when characterized by a terahertz time-domain spectroscopy (THz-TDS) system. We start this article by describing our design strategy – matching the guiding conditions between the rod-type 2D PCs and the PPWG. A supercell of the hybrid waveguide is then numerically evaluated, allowing us to fabricate waveguides operating within the desired frequency range. Next, we measure the transmission of the fabricated waveguides exploiting the THz-TDS method to demonstrate the wideband, single-mode, low-loss, low-GVD propagation of terahertz radiation. Our work has a potential to pave the way for compact, multi-functional, chip-like waveguide devices used in future densely integrated terahertz systems.

## 2. Design and Fabrication

### 2.1. Guiding in Hybrid Structure

**Figure 1** presents a schematic of the proposed hybrid terahertz waveguide, where the 2D PCs composed of dielectric pillars in a square lattice arrangement are sandwiched in the PPWG. Since no light can propagate through the pillars in the PBG, we remove a row of pillars (along the *y*-axis in Figure 1) to create a compact air channel. Nevertheless, without metallic plates, the mode is weakly guided as most power leaks out of the defect in the vertical direction (*z*-axis in Figure 1).[25] Hence the PPWG has to be implemented to vertically confine the power. As a result, the hybrid waveguide could provide a complete confinement for wave guidance. Geometrical parameters are denoted in Figure 1, where the plate separation (or equivalently, the pillar height) is *h*, the in-plane lattice constant is *a*, and the pillar diameter is *d*. In order to achieve low loss and high dielectric constant at terahertz frequencies, the PC pillars are made of high-resistivity float-zone silicon (fabricated from 4" single-side-polished <100> silicon wafer of thickness of 300±25 μm and resistivity of 10 kΩ·cm). We select gold for the metallic plates to facilitate eutectic bonding with silicon.

We begin our design with the simple PPWG with a plate separation of $h$. In order to ensure a rigorous single-mode operation of the hybrid waveguide, the cut-off frequency of the second-order transverse-magnetic (TM) mode (TM$_1$) in PPWG should be higher than the entire PBG of the 2D PCs. The cut-off frequency ($f_{c,m}$) of TM$_m$ mode is governed by the plate separation $h$, which can be expressed as $f_{c,m} = mc/2nh$ (where $c$ is the speed of light in vacuum, and $n$ is the refractive index of the medium between the plates, $n = 1$ here).[13] For convenience, in calculating the cut-off frequency the plate dimension is assumed to be infinite in the in-plane direction (*xy*-plane in Figure 1), and the metal is assumed as a perfect electric conductor (PEC). Next, for the 2D PCs (i.e., the square-latticed silicon pillars), it is well known that there is a complete PBG between the first and second bands for TM (the electric vector perpendicular to the *xy*-plane in Figure 1) modes, but no PBG for transverse-electric (TE, the electric vector in the *xy*-plane in Figure 1) modes.[33] Therefore, this type of 2D PCs acts as a reflector for TM modes in the lateral directions (*x*-axis in Figure 1). In addition, the PBG of TM modes can be manipulated by adjusting the structural parameters $a$ and $d$.

Integrating the metallic PPWG and dielectric 2D PCs, we achieve the hybrid terahertz waveguide. To initially determine the structural parameters, we simulate a 2D PC unit cell employing a finite element method based commercial software COMSOL Multiphysics.[34] Floquet periodic conditions are applied to the exterior boundaries of the unit cell to mimic the response of an infinite array in the *xy*-plane. Because of the non-dispersive nature and the high transparency (absorption coefficient of 0.013 cm$^{-1}$) of high-resistivity silicon covering 0.2-1.0 THz, the refractive index of silicon is simply considered as a constant real value of 3.418.[35] Considering a target transmission window at around 0.4 THz, simulations show that, for a fixed $a = 300$ μm, the PBGs of TM mode are 0.334-0.472 THz and 0.284-0.420 THz when $d/a = 0.3$ and $d/a = 0.4$, respectively. To satisfy the criteria that $f_{c,1}$ of PPWG should be no less than the upper-frequency boundary of PBG, here, we consider the plate separation $h = a$ which corresponds to $f_{c,1} = 0.5$ THz.

In order to accurately evaluate the projected band diagram of the guided quasi-TEM mode, we implement a 3D full-wave supercell simulation of the hybrid waveguide. The 3D model comprising 10 silicon pillars on each side of the middle defect is schematically shown in **Figure 2**a, where $a = 300$ μm and $h = a$. Impedance boundary conditions are applied for the parallel metallic plates, where the material feature for gold is described by the Drude model (plasmon frequency and damping constant are $1.37 \times 10^{16} \text{s}^{-1}$ and $4.07 \times 10^{13} \text{s}^{-1}$, respectively).[36] Meanwhile, Floquet periodic conditions and scattering boundary conditions are exploited for modeling surfaces in *xz*- and *yz*-planes, respectively. Figure 2b shows projected band diagrams for hybrid waveguides with $d/a = 0.3$ (blue curve) and $d/a = 0.4$ (green curve), where solid curves illustrate the simulated dispersion relation between the modal eigenfrequency and the normalized wave vector in the propagation direction ($k_y a/\pi$, where $k_y$ is the modal wave vector) while the dashed curves represent the first and second TM bands. Moreover, we include the light line ($f = ck_y/2\pi$, dark grey dotted line) in Figure 2b. As expected, the projected band diagrams show that the guided modes are inside the PBGs – the terahertz waves can propagate in the hybrid waveguide. Additionally, the transmission window of quasi-TEM mode can be shifted along with the PBG which is essentially tuned by the PC structural parameters (windows of 0.329-0.478 THz and 0.305-0.435 THz when $d/a = 0.3$ and $d/a = 0.4$, respectively). This characteristic confirms the scalability of the hybrid waveguide working at other targeted terahertz frequencies. Simulations also indicate the *k<sub>y</sub>* is above the wave vector in free space (see the light line in Figure 2b). Although the mode is a fast wave, it can be still guided due to the vertical confinement provided by the PPWG. In the context of waveguide fabrication, another important parameter that dominates the lateral confinement performance of 2D PCs is the pillar numbers on each side of the middle defect. We simulate modal propagation losses (which reflect the lateral confinement) of the supercells with various pillar numbers. The results suggest that, in order to avoid power leakage in the lateral direction, the PC pillar number on each side of the middle defect should be no less than 6 and 5 for waveguides with $d/a = 0.3$ and $d/a = 0.4$, respectively (more simulation information in Section S1.1, Supporting Information). Furthermore, we estimate the GVDs of the proposed hybrid waveguide and the silicon PC waveguide using the supercell models. Numerical simulations show that, compared to the silicon PC waveguide, the hybrid waveguide has a flatter GVD curve with lower GVD values in magnitude within a wider transmission window (see Figure S2, Supporting Information).

## 2.2. Fabrication

Here, we fabricate the metal/dielectric hybrid waveguide with target parameters, i.e., $a = h = 300$ μm and $0.3 < d/a < 0.4$, according to the design and simulations in the previous section (the target transmission window is at around 0.4 THz). Cutting-edge microfabrication techniques involving photolithography, eutectic bonding, and dry reactive ion etching are employed to fabricate the substrate-free waveguide (more information of fabrication details in Section 5). We show microscope images of the top view and side view of the waveguide samples before bonding to the top gold plate in **Figure 3**a and 3b, respectively, showing the silicon pillars are uniformly arranged in a square lattice. We carry out a supercell simulation using parameters from the fabricated samples (measured values are $a = h = 297$ μm and $d = 0.367a$). Figure 3c shows that the simulated guided mode range of the fabricated waveguide is from 0.314 THz to 0.456 THz (in between those of the waveguides with $d/a = 0.3$ and $d/a = 0.4$, see Figure 2b), illustrating that the fabricated samples satisfy our design specification. Fabricated straight waveguide samples with different lengths of 14 mm, 42 mm and 72 mm are presented in Figure 3d. Each sample has two separate waveguide channels (cyan dashed lines in Figure 3d) with 10 (denoted as Ch-1) and 15 pillars (denoted as Ch-2) on each side of the middle defect, which are indicated in Figure 3e – an inclined-side view of the 42 mm sample. From measured geometrical parameters above, we observe that the air channel dimension is 485 µm × 297 µm, which is at the wavelength scale (~750 µm). As shown at the top of Figure 3d, silicon pillars and one plate of 14 mm waveguide Ch-2 is unfortunately damaged (represented as a broken line). Furthermore, we fabricate an additional waveguide including two 90-degree bends by arranging the pillar distribution (see Figure S4, Supporting Information), which is expected to be able to control the flow of terahertz waves.

## 3. Experimental Results

We characterize all fabricated hybrid waveguide samples using a commercial THz-TDS system (Menlo TeraSmart).[37] In the system, the electric field of terahertz pulses emitted from the terahertz source is linearly polarized along the *z*-direction (see Figure 1), then the free-space-propagating terahertz pulse is coupled into the waveguide using a pair of symmetric-pass lenses (more information of measurement technique and setup in Section S3.1, Supporting Information). In order to record sufficient pulse information in the time domain, the signals for 14 mm, 42 mm, and 72 mm waveguides are measured over 200 ps, 300 ps, and 450 ps, respectively. In the following text, we first measure the waveguide transmissions and raster-scan the waveguide output. Next, we characterize the propagation loss and GVD of the fabricated waveguides exploiting the cut-back method.

### 3.1. Transmission Measurement

First, we perform the transmission measurement of different waveguides and channels in **Figure** 4, which are normalized with the reference signals without waveguide but with the symmetric-pass lenses positioned confocally and a pinhole at the focal point (the pinhole is employed for waveguide transmission measurement, see more information in Section S3.1, Supporting Information). The 3-dB bandwidth is estimated via averaging the oscillations of high transmittance from the spectrum, which is indicated as the shaded region. Figure 4a and 4b show the transmittance for Ch-1 (comparison between 14 mm and 42 mm samples) and Ch-2 (comparison between 42 mm and 72 mm samples), respectively. As expected, increasing the waveguide length results in less transmittance and narrower transmission window. It is noticeable that the 14 mm waveguide Ch-1 offers a broad bandwidth beyond 100 GHz. Moreover, an important parameter in the hybrid waveguide design is the pillar numbers on each side of the middle defect, which dominates the lateral confinement loss (previously investigated via simulations, see Section S1.1, Supporting Information). Figure 4c compares the measured transmission performance between Ch-1 (10 pillars on each side of the channel) and Ch-2 (15 pillars on each side of the channel) of the 42 mm sample. We can attribute the slight variations of transmission bandwidth to fabrication tolerance, as the waveguides with 10 and 15 pillars on each side of the middle defect have nearly identical loss curves from supercell simulations (see Figure S1, Supporting Information). To further enhance the transmission bandwidth of the hybrid waveguide, a possible solution is increasing the middle defect width. Simulations show that the low-loss window (below 0.05 dB $mm^{-1}$) could be up to 138 GHz (almost fully utilizing the PBG, see Figure 3c) for waveguide with increasing defect width by 100 μm (total width of 585 μm, more simulation results in Section S1.3, Supporting Information). The enhanced low-loss (below 0.05 dB $mm^{-1}$) bandwidth of the 585 μm waveguide is 3.14 times wider than that of the fabricated one. In addition, the coupling efficiency of the proposed hybrid waveguides can be further improved by either employing a properly designed horn antenna to be assembled at the waveguide input and output ends, or integrating a source or detector into the waveguide directly.

Second, to verify the transmitted mode is indeed guided in the channel (i.e., air defect), we image the distributions of the *y*-component of power ($P_y$, i.e., the power along the waveguide propagation direction). The fabricated samples are mounted on a two-axis stage with a scan step of 50 μm in *xz*-plane (see Figure S5, Supporting Information). The image at a discrete frequency can then be extracted from the scanned data containing the entire spectral information for each data pixel. We should note that this scanning method yields images that are affected by the square of the coupling coefficient to the waveguide, as both the input and output are translated with respect to the center of the waveguide in the raster scan. Here, we show experimental normalized $P_y$ images of the 14 mm waveguide Ch-1 at frequencies in (0.37 THz) and out of (0.19 THz) the transmission window in **Figure 5**a and 5b, respectively. For comparison, Figure 5c shows the simulated distribution of $P_y$ of guided mode at 0.37 THz. Comparing Figure 5a and 5c at the same 0.37 THz, both experimental and simulated results illustrate that the terahertz radiation can be well confined in the wavelength-scale channel within the transmission window. Note that the simulated $P_y$ is extracted from the supercell simulation (inside the waveguide), while the experimental image is measured at the end of the waveguide (out of the waveguide). This explains the difference observed between the experimental (Figure 5a) and simulated (Figure 5c) $P_y$ distributions. On the contrary, at 0.19 THz, the $P_y$ of 14 mm waveguide Ch-1 largely spreads in the PC pillar region, where the terahertz wave is hardly guided. The experimental images of $P_y$ of all the fabricated straight waveguides (see images of the remaining waveguides in Figure S9, Supporting Information) further confirm that terahertz waves are guided in the compact wavelength-scale channels and completely confined by the hybrid structure.

We have also measured the transmission spectra of the 72 mm waveguide Ch-1 and the 90-degree bend waveguide. For the 72 mm waveguide Ch-1, we observe a sharp transmission dip at 0.383 THz with a 19-dB extinction ratio and a 15-GHz full-width half-maximum (see Figure S6, Supporting Information). One possible reason could be fabrication tolerances – one or several pillars may be missing, making it a 2D PC cavity-like structure.[26] We point this out as a small rectangle in Figure 3d. The experimental measurement of 90-degree bend waveguide shows a transmission window of 0.318-0.402 THz, demonstrating that the terahertz radiation can propagate along with an assigned path (see Figure S7, Supporting

Information). The bend induced scattering loss is estimated using 3D full-vector numerical simulations. By comparing the propagation losses among waveguide structures with 1-bend defect, 2-bend defect, and straight defect, we observe the scattering loss is around 0.25 dB bend$^{-1}$ (more simulation information in Section S3.2, Supporting Information).

**3.2. Waveguide Characteristics**

Here, we investigate the transmission properties of all the intact hybrid PC waveguide samples using a THz-TDS system to validate the theoretical results. We adopt the cut-back approach (i.e., comparing waveguides with different lengths) for loss and GVD measurements.[5] To conduct a sufficiently precise characterization, the components used in the THz-TDS system are adjusted for the best possible coupling performance, and the waveguide signal is captured at the center of the air channel based on waveguide imaging. Since the Ch-1 and Ch-2 have almost similar transmission bandwidth and loss (validated both theoretically and experimentally), we have five sets of measurements from waveguides with different lengths to extract the loss and GVD characteristics.

The propagation loss can be obtained by comparing the measured intensities of two signals. As shown in **Figure 6**, the red solid curve shows the average value of the experimental losses from five data sets, and the error bars indicate the standard deviation of the measurements. For a direct quantitative comparison, we include the theoretical loss (blue solid curve) in Figure 6, which is calculated from the numerical supercell simulation of the fabricated waveguide. The green and orange shaded regions present the experimental bandwidths for losses below 0.1 dB mm$^{-1}$ and 0.05 dB mm$^{-1}$, respectively. In comparison with the theoretical results, the experimental characterization indicates a narrower low-loss transmission band, which is restricted by the measured transmission windows of long waveguides (42 mm and 72 mm samples, see Figure 4). Nonetheless, within the frequency range for measured losses below 0.05 dB mm$^{-1}$ (0.367-0.411 THz), we observe a good agreement between experimental and theoretical results. The measured minimum losses are down to 0.01 dB mm$^{-1}$ at frequencies around 0.37-0.39 THz. **Table 1** summarizes the loss and the single-mode transmission bandwidth of our hybrid waveguide as well as the existing all-metallic and all-dielectric terahertz PC chips.[16, 19] Compared to the all-silicon PC waveguide, our hybrid waveguide provides an over doubled single-mode operating bandwidth– the characterized low-loss bandwidth is 44 GHz with losses no higher than 0.05 dB mm$^{-1}$. According to the latest IEEE 802.15.3d-2017 standard for high data rate (up to 100 Gb s$^{-1}$) wireless communication at terahertz frequencies,[38] the proposed hybrid waveguide is able to support the first six or the all eight bandwidths under the loss criterion of 0.05 dB mm$^{-1}$ or 0.1 dB mm$^{-1}$, respectively; while the all-silicon PC waveguide could cover the first two bandwidths (see summarized experimental low-loss bandwidths in Table 1). It should be noted that structural parameters of the hybrid waveguide need updates to satisfy the frequency range of the IEEE standard (0.252-0.325 THz). For the all-metallic PC chip, although its broad transmission bandwidth (up to 500 GHz) satisfies all IEEE standard suggested bandwidths, our hybrid waveguide provides two times lower losses thanks to the low material absorption of silicon in terahertz frequencies. It should be noted that the narrower bandwidth of the hybrid waveguide in contrast to that of all-metallic one is fundamentally limited by the dielectric PBG). Furthermore, we expect that, by optimizing the middle defect width (see discussions in Section S1.3, Supporting Information) and improving the fabrication technique, for instance avoiding the residues at the bottom of silicon pillars (see Figure 3b), the single-mode bandwidth of the hybrid waveguide could be more than 100 GHz for losses below 0.05 dB mm$^{-1}$.

We then characterize the GVD of the hybrid waveguide. The GVD is calculated as $GVD = \partial^2\beta/\partial\omega^2$, where $\beta$ and $\omega$ are the frequency-dependent propagation constant of the guided wave (determined by the phase information of measured signals) and the angular frequency, respectively. The experimental GVD data of each waveguide are first fitted by a 7-th order polynomial function within the transmission window for losses below 0.1 dB mm$^{-1}$ (see Figure 6), followed by an average of the five waveguides. The experimental and theoretical GVD values are shown as red and blue solid curves in **Figure 7**, respectively, where the error bars illustrate the standard deviation of the measurements. We notice that the GVD values are less accurate near to the PBG edges where the signal to noise ratio decreases and standard deviations increase. The experimental GVD curve is in good agreement with the simulated one, featuring negative and positive GVD values at low and high frequencies within the transmission window, respectively [zero GVD is at 0.413 THz, see the crossing between the red solid curve and dark grey dashed line ($GVD = 0$) in Figure 7]. Over 0.367-0.411 THz (losses lower than 0.05 dB mm$^{-1}$, see Figure 6), the GVD values are all negative, from -8.4 ps THz$^{-1}$ mm$^{-1}$ to -0.8 ps THz$^{-1}$ mm$^{-1}$, illustrating a relatively flat, near-zero dispersion curve, thanks to the quasi-TEM mode guidance. Comparing with the silicon PC waveguide, the proposed hybrid waveguide holds advantages of lower GVD covering a wider bandwidth (see Figure S2, Supporting Information). Nonetheless, it should be noted that the GVDs of all-metallic PC waveguides would be further reduced over a wider transmission window (see Table 1) in contrast to the hybrid one, due to their wider bandgap.

## 4. Conclusion

In summary, we have developed a compact, substrate-free terahertz hybrid waveguide chip consisting of silicon PCs sandwiched in parallel gold plates. Our approach provides an excellent balanced design – comparing with all-dielectric PC waveguide, the proposed hybrid waveguide has an improvement of more than twice wider single-mode bandwidth with lower GVD; meanwhile, the propagation losses of our solution are two times lower than that of all-metallic PC waveguides. Experimental characterizations demonstrate the fabricated straight waveguides propagate terahertz radiation with losses down to 0.05 dB mm$^{-1}$ and near-zero GVD values (from -8.4 ps THz$^{-1}$ mm$^{-1}$ to -0.8 ps THz$^{-1}$ mm$^{-1}$) covering a single-mode regime (0.367-0.411 THz, bandwidth of 44 GHz). Such a wide spectral window supports the first six bandwidths (even the all eight bandwidths but with higher transmission losses of 0.1 dB mm$^{-1}$) defined by the IEEE 802.15.3d-2017 standard. Measurements also show that the hybrid waveguides with modified pillar arrangements are able to act as 90-degree bends and a bandstop filter, illustrating that the advanced PC based functionalities in optics can be readily transferred to the terahertz field. Future terahertz devices based on the proposed waveguide platform, including filters, splitters, multiplexers, and antennas, can be achieved using the proposed multi-wafer silicon-based microfabrication approach. More interestingly, we can anticipate an integrated terahertz hybrid chip incorporating multiple functional devices, even implementing tunable and active components, which itself is an integrated system within a single photonic chip. In a nutshell, we envision that this compact terahertz chip would play a significant role in future dense-integrated broadband terahertz systems applied for communications, sensing and other applications.

## 5. Experimental Section

*Fabrication Technique*: Here, we describe the detailed fabrication procedure of the hybrid waveguide. First, a 4" high-resistivity silicon wafer (single-side-polished of <100> orientation) of thickness 300±25 μm and resistivity 10 kΩ·cm, and two standard silicon wafers (to be used as the supporting substrates, wafer A is 4", and wafer B is 3") are deposited with 200 nm of gold using an electron beam evaporator (PVD75, Kurt J. Lesker), as shown in **Figure 8**a. Next, eutectic bonding (Figure 8b) is conducted to bond the gold-coated high-resistivity silicon wafer to the standard silicon wafer A, using a wafer bonder (SB6 SUSS Microtech). Bonding is performed at 400 ºC, a tool pressure of 1000 mbar and a dwelling time of 20 min.[39, 40] After that, the surface of the high resistivity silicon is patterned with AZ4562 photoresist using a chrome mask and a UV mask aligner (SUSS MA6/BA6 Mask Aligner), as shown in Figure 8c. This is followed by deep reactive ion etching (Figure 8d) to transfer the desired pattern onto the high-resistivity silicon wafer with octafluorocyclobutane ($C_4F_8$) and sulfur hexafluoride ($SF_6$) as etchant gases (Estrelas-Oxford Plasma Pro 100). Prior to the dicing of the wafer into individual device pieces, the wafer is first coated with a layer of AZ5214E photoresist and baked at 100 °C for 10 min to protect the high-resistivity silicon pillars during dicing (Figure 8e). The etched wafer is then cut into individual device pieces using a laser micromachining equipment (dicing saw-DISCO DAD 321- NanoFab), as presented in Figure 8e. Similarly, the unused gold-coated silicon wafer B is also diced into the same device dimension (Figure 8e). Last, each device piece and its corresponding standard silicon piece are bonded together to realize the proposed hybrid structure (Figure 8f). The fabrication technique enables achieving compact, substrate-free, chip-like waveguide structure, which is an excellent platform for the next generation of terahertz devices and holds huge potential for dense integration of terahertz systems.

**Supporting Information**
Supporting Information is available from the Wiley Online Library or from the author.

**Acknowledgements**
H. Li acknowledges support from the National Key R&D Program of China No. 2018YFB1801003, the National Natural Science Foundation of China No. 61805010, and the Beijing Natural Science Foundation No. 4192048. S. Atakaramians acknowledges support from the Australian Research Council (ARC) under the Discovery Early Career Project Award No. DE140100614, and the UNSW Sydney Scientia Fellowship. This work was performed in part at the Micro Nano Research Facility at RMIT University in the Victorian Node of the Australian National Fabrication Facility (ANFF).

**Conflict of Interest**
The authors declare no conflict of interest.

Received: ((will be filled in by the editorial staff))
Revised: ((will be filled in by the editorial staff))
Published online: ((will be filled in by the editorial staff))

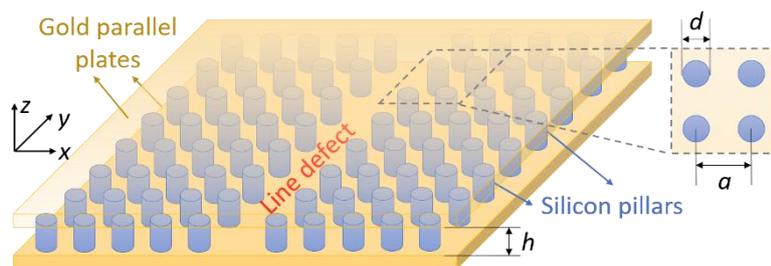

**Figure 1.** Schematic of the hybrid waveguide consisting of silicon PC pillars and gold parallel plates. Inset is a top view of PCs with defined parameters.

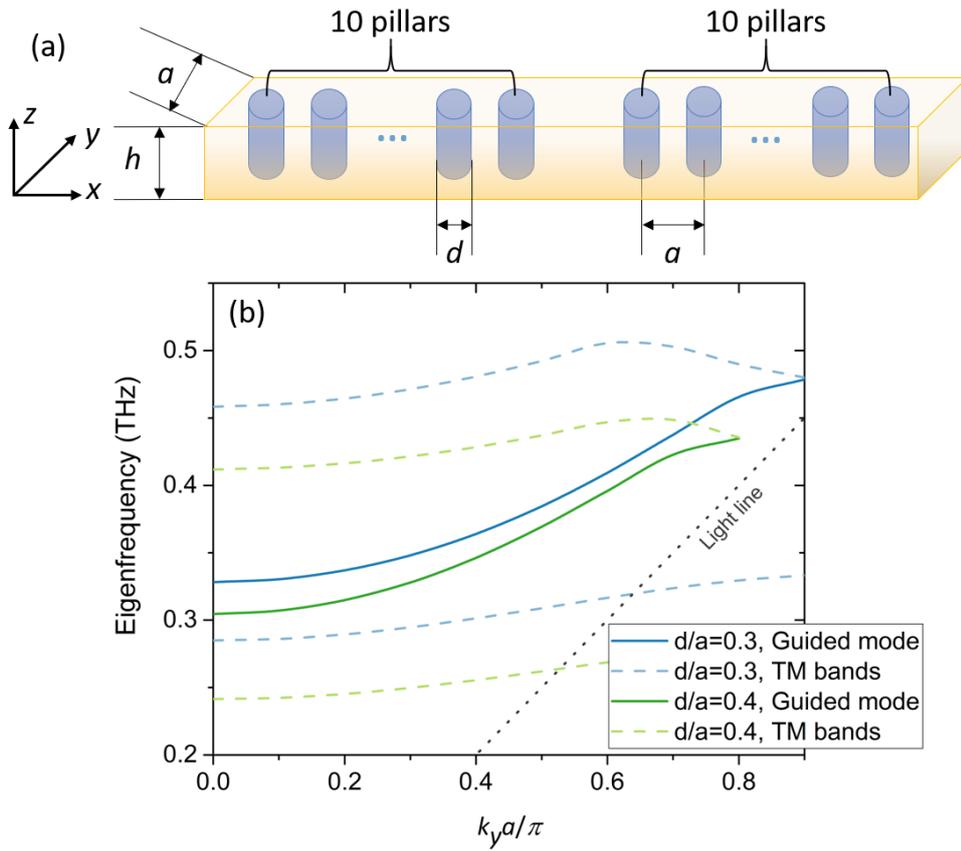

**Figure 2.** Supercell simulations of the hybrid waveguide: a) Schematic of the numerical model with its geometrical parameters. b) Projected band diagrams for TM modes of the hybrid waveguides with $d/a = 0.3$ (blue) and $d/a = 0.4$ (green). Dark grey dotted line represents the light line.

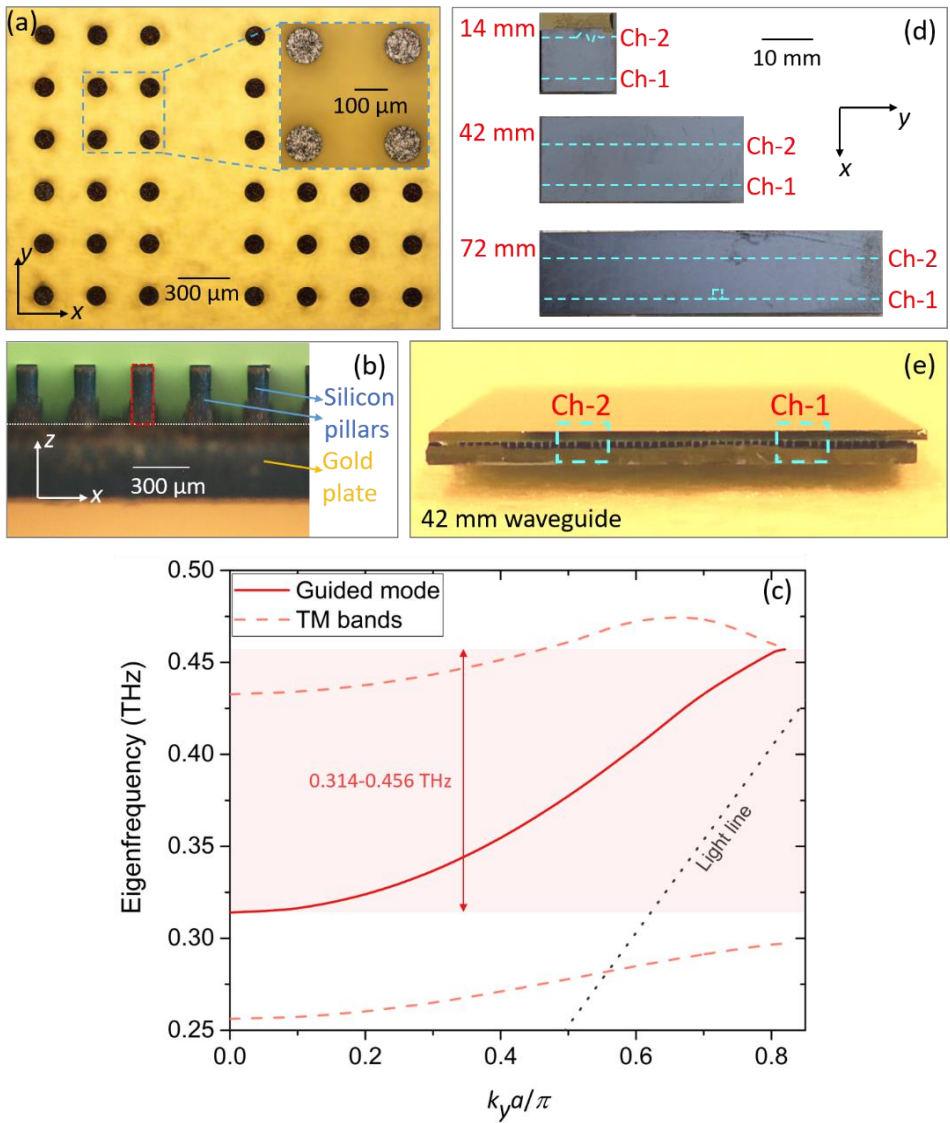

**Figure 3.** Fabricated waveguide samples and corresponding numerical simulation results: a) Microscope image of the top view of the sample before bonding to the top gold plate. The inset shows a section of the magnified structure. b) Microscope image of the side view of the same sample, where the red-dashed rectangle highlights a silicon pillar and the white dotted line represents the boundary between silicon pillars and one gold plate. c) Projected band diagrams for the fabricated waveguide. d) Top view of the fabricated samples with lengths of 14 mm, 42 mm and 72 mm. e) Inclined-side view of the 42 mm sample.

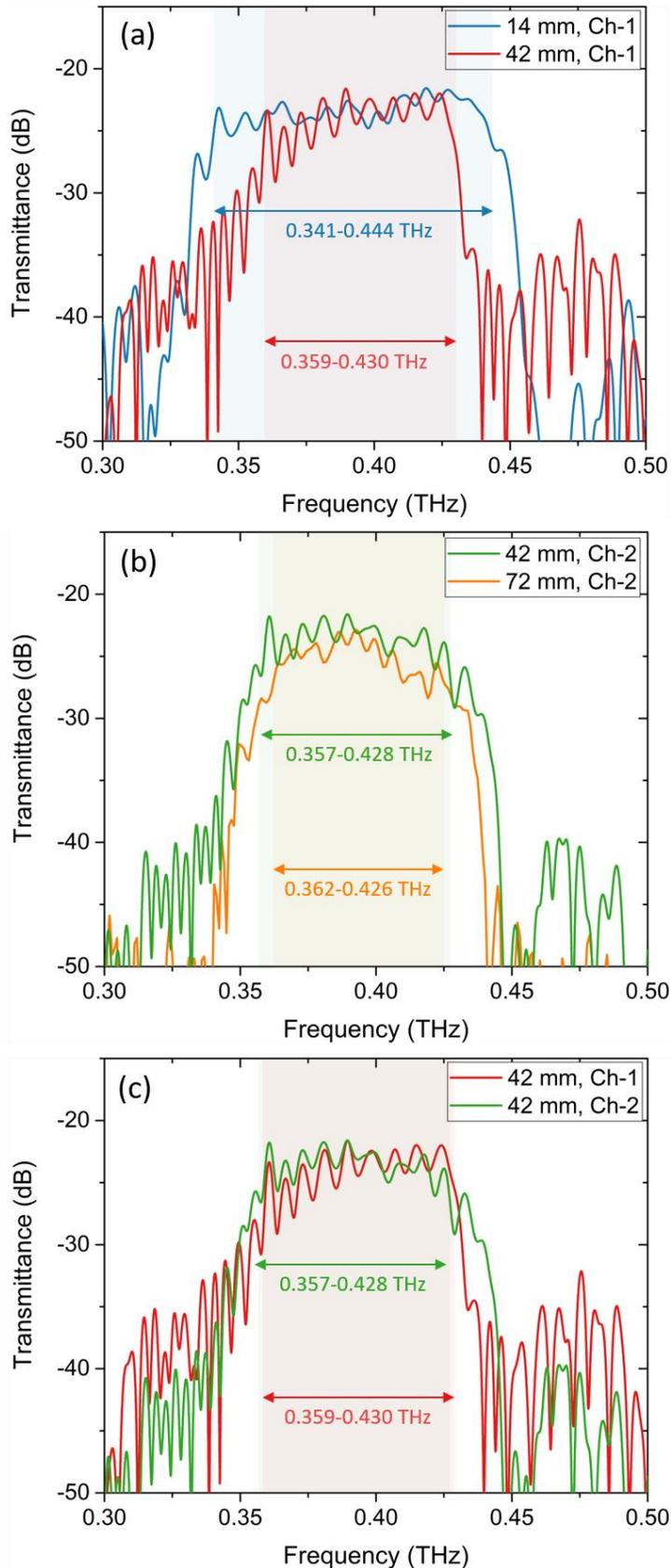

**Figure 4.** Normalized transmission spectra of the fabricated waveguides: a) Comparison of Ch-1 between 14 mm (blue solid curve) and 42 mm (red solid curve) samples. b) Comparison of Ch-2 between 42 mm (green solid curve) and 72 mm (orange solid curve) samples. c) Comparison between Ch-1 (red solid curve) and Ch-2 (green solid curve) of the 42 mm sample.

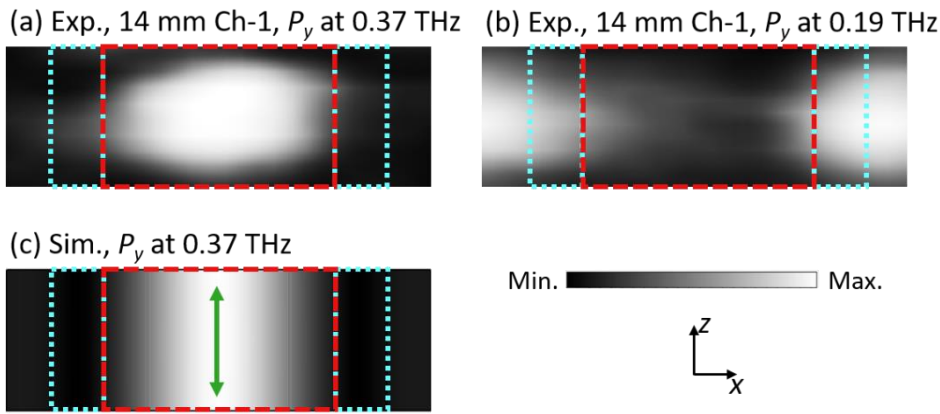

**Figure 5.** Normalized $P_y$ distributions. Scanned $P_y$ distributions of 14 mm waveguide Ch-1 at a) 0.37 THz and b) 0.19 THz. c) Simulated $P_y$ distribution at 0.37 THz. The red dashed rectangle, cyan dotted rectangle, and green arrow represent the channel (size of 485 μm × 297 μm), the adjacent pillars, and the simulated polarization direction of the electric field, respectively.

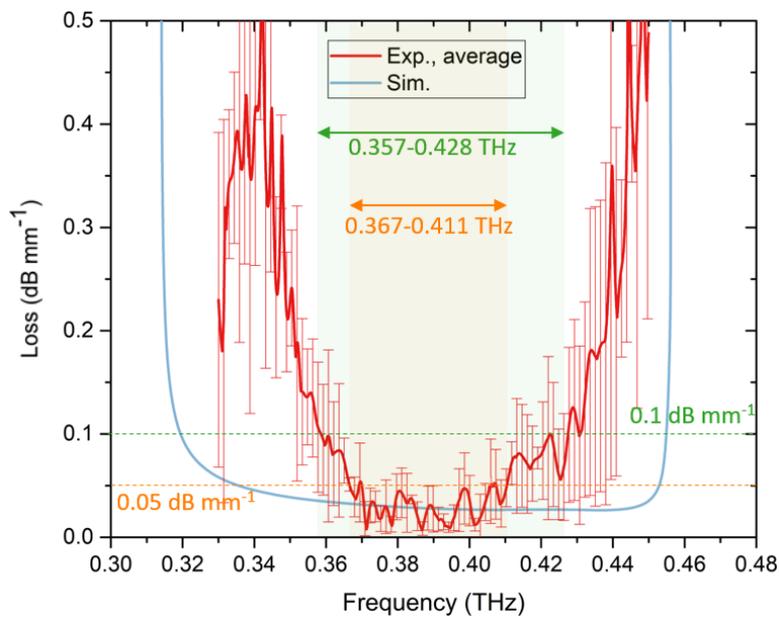

**Figure 6.** Waveguide propagation losses as a function of frequency. Red and blue solid curves show experimental and theoretical results, respectively. The error bars indicate the standard deviation of the measurements.

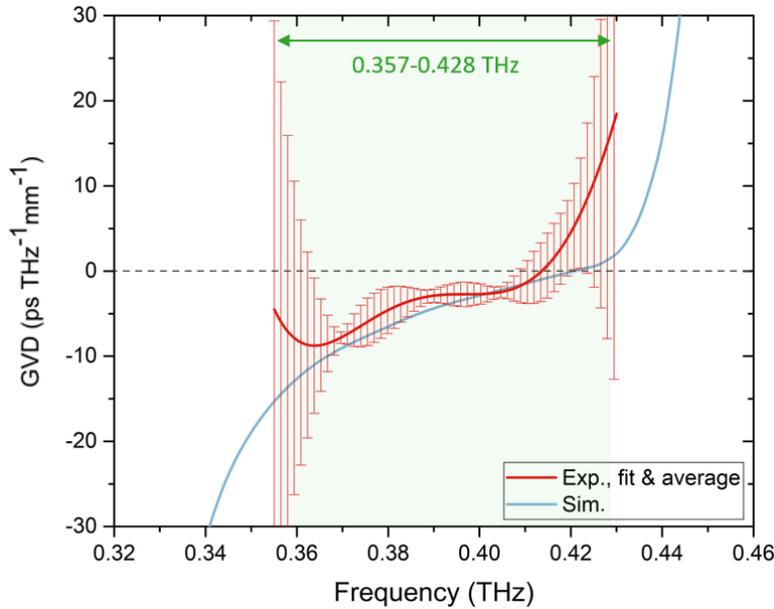

**Figure 7.** Waveguide GVD values as a function of frequency. Red and blue solid curves show experimental and theoretical results, respectively. Dark grey dashed line represents $GVD = 0$. The error bars indicate the standard deviation of the measurements.

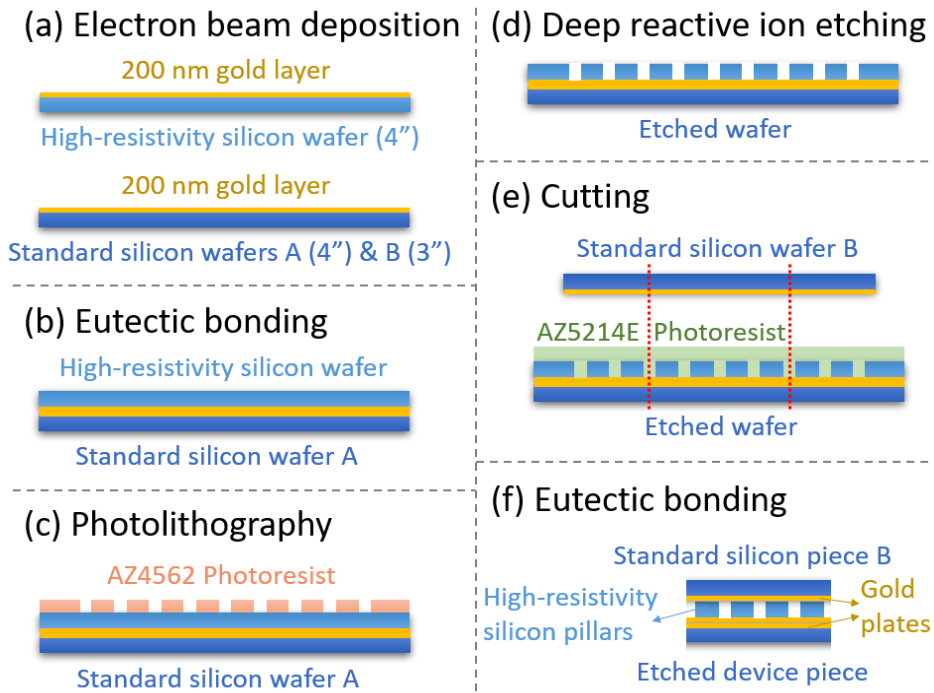

**Figure 8.** Schematical fabrication procedure of the hybrid waveguide: a) Electron beam deposition of gold on silicon. b) Eutectic bonding between the gold-coated high-resistivity silicon wafer and the gold-coated standard silicon wafer A. c) Standard photolithography. d) The wafer after deep reactive ion etching. e) Cutting of the etched wafer and the gold-coated standard silicon wafer B. f) Eutectic bonding between the etched device piece and the gold-coated standard silicon piece B.

**Table 1.** Losses and single-mode transmission bandwidths of the PCs based terahertz waveguides.

| Waveguide | Loss [dB mm$^{-1}$] | Bandwidth [GHz] | Reference |
|---|---|---|---|
| Hybrid (gold/10 kΩ·cm silicon) waveguide | <0.1 (exp.) <br> <0.05 (exp.) | 71 (exp.) <br> 44 (exp.) | - |
| Metallic (gold) waveguide | 0.1-0.19 (exp.) | 500 (exp.) | [16] |
| Dielectric (20 kΩ·cm silicon) waveguide | <0.01 (sim.) <br> <0.01 (exp.) | 18 (sim.) <br> 5 (exp.) | [19] |